\documentclass{pasa}
\usepackage{enumerate}
\usepackage{graphicx}
\usepackage{multicol}
\usepackage{lscape}
\usepackage{longtable}
\usepackage{rotating}
\usepackage{pdflscape}

\title[Verifying timestamps of occultation observation systems]{Verifying timestamps of occultation observation systems}
\author[Barry et al.]{M.A.(Tony) Barry$^{1}$, Dave Gault$^{2}$, Greg Bolt$^3$, Alistair McEwan$^1$, Miroslav D. Filipovi\' c$^{4}$ \and Graeme L. White$^4$\\
\affil{$^1$University of Sydney, Electrical and Information Engineering Department, Camperdown, NSW 2006 Australia}
\affil{$^2$Kuriwa Observatory (MPC E28) 22 Booker Rd, Hawkesbury Heights, NSW 2777}%
\affil{$^3$Craigie Observatory (MPC 321), 295 Camberwarra Drive, Craigie, WA 6025, Australia}
\affil{$^4$University of Western Sydney, Locked Bag 1797, Penrith South, DC, NSW 1797, Australia}}%

\jid{PASA}
\doi{10.1017/pas.\the\year.xxx}
\jyear{\the\year}

\begin{document}%
\begin{abstract}
We describe an image timestamp verification system to determine the exposure timing characteristics and continuity of images made by an imaging camera and recorder, with reference to Coordinated Universal Time (UTC). The original use was to verify the timestamps of stellar occultation recording systems, but the system is applicable to lunar flashes, planetary transits, sprite recording, or any area where reliable timestamps are required. The system offers good temporal resolution (down to 2~msec, referred to UTC) and provides exposure duration and interframe dead time information. The system uses inexpensive, off-the-shelf components, requires minimal assembly and requires no high-voltage components or connections. We also describe an application to load FITS (and other format) image files, which can decode the verification image timestamp. Source code, wiring diagrams and built applications are provided to aid the construction and use of the device.
\end{abstract}
\maketitle%

\section{INTRODUCTION }
 \label{sec:intro}
The occultation of distant stars by solar system bodies (asteroids, dwarf planets, TNOs, etc) provides a method to characterise the nature of the solar system bodies to a resolution that cannot be matched except by space probe observations \cite{PHOTpaper}.
	
An occultation recording consists of an earth station observing the star and asteroid coalescence and monitoring the light output over time (the light curve of the occultation). As the asteroid occults the star, the light flux is reduced. The recording aims to capture the time (UTC) when the light flux changes and the manner in which it changes to determine a chord through the body. The recording also allows detection of the presence (if any) of an atmosphere, and satellites or ring structures \cite{SolarSystemOccPaper}.

With several earth stations observing the same event, a series of adjacent chords can be drawn, providing more information about the asteroid and environs. The diameter of the parent body can be more precisely estimated, the body shape can be examined for oblateness, and any satellites or ring structures can have their orbits determined\cite{QuaoarOccPaper}.

All these measurements depend on the time stamp of each image in the occultation recording being referenced to a known time standard such as UTC. Accuracy of timebase should be to within a millisecond \cite{PHOTpaper}.

\section{MOTIVATION}
 \label{sec:motivation}
In the case of a recent occultation of 10199~Chariklo, a $\sim$250~km diameter member of the Centaur group orbiting between Saturn and Uranus, there were fourteen observing stations, spread across more than 1000~km of South America, of which eight observed the occultation \cite{CharikloNaturePaper}.

The occultation was remarkable because it was the first observation of two rings, of 7 and 3~km width, orbiting the primary body at a distance of 391 and 405~km.

Unfortunately, there were disparities in absolute time consensus between two of the observing stations, housed in the same observatory, of 1.6~seconds. 

The shadow transit speed of the occultation was calculated to be 21.6~km~sec$^{-1}$, and so this disparity represents a disjunction of about 35~km in measurements from two side-by-side stations. The measurements were able to be adjusted because the two systems were side-by-side and observed the same event, and previous observations indicated one system had a record of temporal fidelity while the other was known to have unexplained offsets up to 2.5~seconds from true.

There were also timing disparities found with one other station which also observed the occultation, but it could not be adjusted for, because the offset was not able to be characterised. Consequently, the information from this station was not used in the data reduction for the observation of the rings of Chariklo \cite{CharikloNaturePaperSupplementaryInfo}.

\section{CURRENT PRACTICE FOR OCCULTATION TIMING SYSTEMS}
 \label{sec:currentPractice}

Most occultation systems in use today rely on either Global Positioning Systems (GPS) based time sources for fidelity to UTC, or use Network Time Protocol (NTP) as the method to synchronise the imaging system computer clock with a Stratum~1 timeserver through a link to the internet \cite{QuaoarOccPaper}. Previous methods of timestamping using the reception of specialised radio broadcasts such as Radio WWV in the Americas or Radio VNC in Australia are no longer available or soon to be phased out \cite{MiroslavPaper}.

The aim for accuracy of the timestamp is to be within a millisecond of true, and several GPS based devices exist for the purpose of time-stamping analog video (CVBS), including those devised by BlackBox Camera, Alexander Meier Elektronik, PFD Systems and IOTA.

Specialised digital occultation camera systems such as PHOT, PICO, POETS and MORIS trigger their acquisitions based on GPS signals stated to be within a millisecond of UTC \cite{PHOTpaper} \cite{PICOpaper} \cite{POETSpaper} \cite{MORISpaper}.

Camera systems originally intended for astrophotography generally use the NTP based PC system clock to provide the header timestamp \cite{CharonOcc2008Paper}.

The time stamp can be either written onto the image itself (in the case of analogue video) or embedded in the image header (for digital video or FITS files).

\section{VERIFICATION OF TIMESTAMP FIDELITY}
 \label{sec:verification}
 
The duration of a camera exposure has previously been verified in popular literature \cite{DavidhazyPaper}, by imaging the raster of an analog video screen (cathode ray tube). The timebase of a raster is well controlled, and for short exposures provides an elegant solution. Counting raster lines represented the basis for broadcast camera accreditation before digital cameras became common. Unfortunately, CRT displays are becoming rare, and connecting to the timebase of the display to provide UTC synchronisation requires high voltage interfacing.

Time stamp fidelity to UTC has previously been verified to field resolution (16.7~msec for NTSC, 20~msec for PAL). This is done by observing an optical event such as a flashing light emitting diode (LED) whose time of illumination is well established from electrical measurements and fiducial sources such as the 1PPS signal from Global Positioning System (GPS) receivers \cite{HarringtonPaper}. While this does not establish the duration of the image, it does give an upper limit for the timestamp of the particular frame where the 1PPS LED was observed.

\section{SOLUTION FOR VERIFYING TIMESTAMPS}
 \label{sec:solutionDescription}

One way to verify an image timestamp is to provide an optical device which is crafted to indicate the passage of time in an unambiguous way. When a camera under test takes an image of the device, the image contains information which can be decoded to produce an image start and image stop time.

We describe such a device in this paper. It consists of an array of 500 LEDs, of which only one LED is illuminated at any one time, and only for a short time (e.g.~2~msec). The array begins its first (top-left) LED illumination at a UTC integer second boundary and over the course of that second illuminates each LED for 2~msec, one after the other down the first column, then down the subsequent columns to the right. 

This is therefore a moving dot of light, and the camera system being verified records the moving dot of light. In each image, some of the LEDs are illuminated due to the camera recording during the time when those LEDs were active, while others are dark, and the position of the illuminated LEDs in each image provides an unambiguous optical timestamp. 

The device uses an internal GPS receiver for reference to UTC, and as per good metrology practice, has a timebase which is accurate to better than a tenth of the basic unit of measure (i.e. for a 2~msec measure, the accuracy should be $<$0.2~msec).

The time period from illumination of the first LED to the last is the sweep time. The device provides 4 sweep time settings; 1~second, 2~seconds, 5~seconds and 10~seconds. In this paper we describe the results for the 1~second sweep as this has the most rigorous timing requirements, and use the 1~second, 2~second, and the 5~second sweep as a basis for testing two different cameras in Sections~\ref{sec:test1}~and~\ref{sec:test2}. 

We also describe an image analysis program for PC, Mac, and Linux, which can decode the clock display. 

Source code, wiring diagrams and built applications are provided to aid the construction and use of the system. 

The present verification system has been named SEXTA (Southern EXposure Timing Array) after work by Dangl \cite{DanglWebPage} - see Acknowledgements - and was developed to verify image timestamps for a digital occultation camera and recorder system developed by the present authors \cite{ADVSpaper}.

\section{METHOD}
 \label{sec:method}
\begin{figure}[h!]
\begin{center}
\includegraphics[width=\columnwidth]{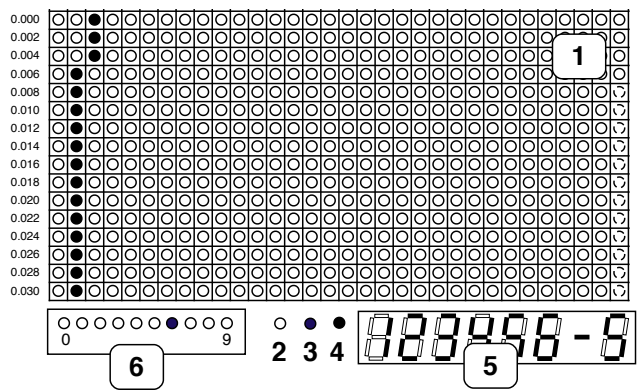}
\caption{SEXTA display. See Section~Method for details. }
 \label{Fig1}
\end{center}
\end{figure}

The camera system under test is set up to view the SEXTA panel, as shown in Figure~\ref{Fig1}. The item numbers of this list correspond to Figure~\ref{Fig1} items.

\begin{enumerate}
\item A panel of 500 LEDs. The first LED is illuminated at UT boundary, and each successive LED glows for a set time. For a 1-second sweep, the time is 2~msec. The illumination pattern is down a column, then to the right. 
\item A "1PPS" LED that flashes with the arrival of the 1PPS signal from the GPS. The fiducial point of the 1PPS LED is the off-to-on transition.
\item A "Lock" LED to show the Dot Matrix Display (DMD) panel is locked to GPS;
\item An "Almanac-OK" LED to indicate that the GPS almanac is current;
\item A 7-segment LED array next to the panel of LEDs to indicate UT hours, minutes, and seconds, and the number of satellites in the GPS fix; and 
\item An array of ten LEDs to indicate the last digit of UT integer seconds (0-9).
\end{enumerate}

From Figure~\ref{Fig1}, exposure start time is 12:34:56.038; exposure end time is 12:34:56.070 UT. The system has 5 satellites in view, the almanac is current and the panel is locked to GPS. Because the image does not contain a UT integer boundary, the 1PPS LED is not lit.
	
With the camera and recorder under test we take images of the panel. Each image shows the optical timestamp provided by SEXTA, and is internally timestamped by the camera system using its own method.

\section{TESTING}

A timing analysis of the 1~second sweep was performed, with each LED illumination being measured for duration. The results of the analysis are shown below in Table~\ref{tab1}.

\begin{table}[h!]
\caption{The analysis of timing of the 1 sec sweep.}
\begin{center}
\begin{tabular}{@{}lc@{}}
\hline
TARGET	& 2 msec\\
\hline
MEAN	    & 1.997 msec\\
\hline
MAX	    & 2.003 msec\\
\hline
MIN	    & 1.987 msec\\
\hline
STDEV	& 0.0024 msec\\
\hline\hline
\end{tabular}
\end{center}
\label{tab1}
\end{table}

The target of 2~msec illumination per LED was met, with the error being less than the desired 0.2~msec accuracy. An oscilloscope was connected to the unit to determine the latency between the GPS fiducial signal (the 1PPS) and the illumination of the first LED. This was measured at 0.165~msec. The latency between 1PPS and UT unit seconds LED changeover was measured at 0.036~msec. The latency between 1PPS and the 1PPS LED illumination was measured at 0.007~msec. All of these times are below the 0.2~msec accuracy required.

\section{USAGE}
 \label{sec:usage}
The SEXTA display is placed in the imaging system field of view at focus, and powered up. A Reference Image pattern is displayed for ten seconds, which the imaging system should acquire to aid the reader application with positional information of the LEDs on the panel. 
 
When the GPS acquires a fix, the 1PPS begins flashing, and the 500-LED array engages in its synchronisation process, taking around 3 minutes to acquire lock. The Lock LED is illuminated when the process is complete. The GPS may take some time ($<$~15 minutes) to download a current almanac, which contains the GPS-UT offset; this is necessary to ensure SEXTA is producing a correct time stamp, and the currency of the almanac is indicated by the A-OK LED.

SEXTA is ready when the panel scrolls across at the determined rate, the 1PPS LED flashes every second, and the Almanac-OK LED and Lock LED are illuminated constantly. The 7-segment LED array indicates UT hh:mm:ss and the number of satellites in the fix.

The imaging system then acquires images of the panel at the desired settings. When saved, the images are analysed to determine the congruency of the timestamp as saved by the imaging system, and the SEXTA-delivered optical timestamp contained in the image.

To ease the chore of reading optical timestamps, a reader application has been produced to automate the process. The application reads FITS files (and other common formats) and can extract timestamp and exposure duration information from the FITS header if these are present. 

\begin{figure*}[h!]
\begin{center}
\includegraphics[width=17.5cm]{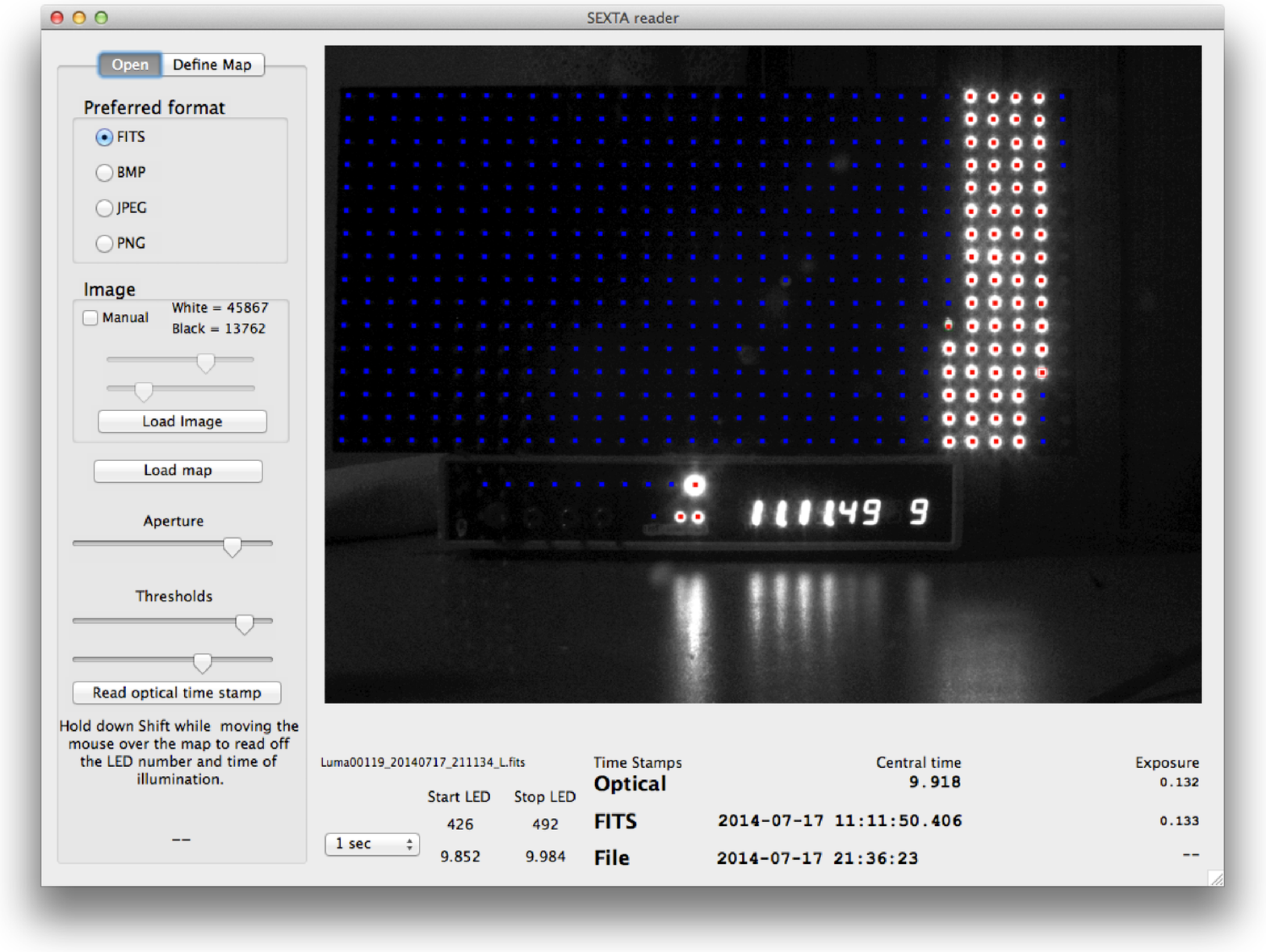}
\caption{SEXTA Reader application screen. For details see Section~\ref{sec:usage}.}
\label{Fig2}
\end{center}
\end{figure*}

In Figure~\ref{Fig2}, the expected position of each LED is shown by blue markers, with red markers where the LED brightness is above the threshold level. The optical, FITS derived and file creation time stamps are compared at the bottom of the window.

\begin{figure*}[h!]
\begin{center}
\includegraphics[width=17.5cm]{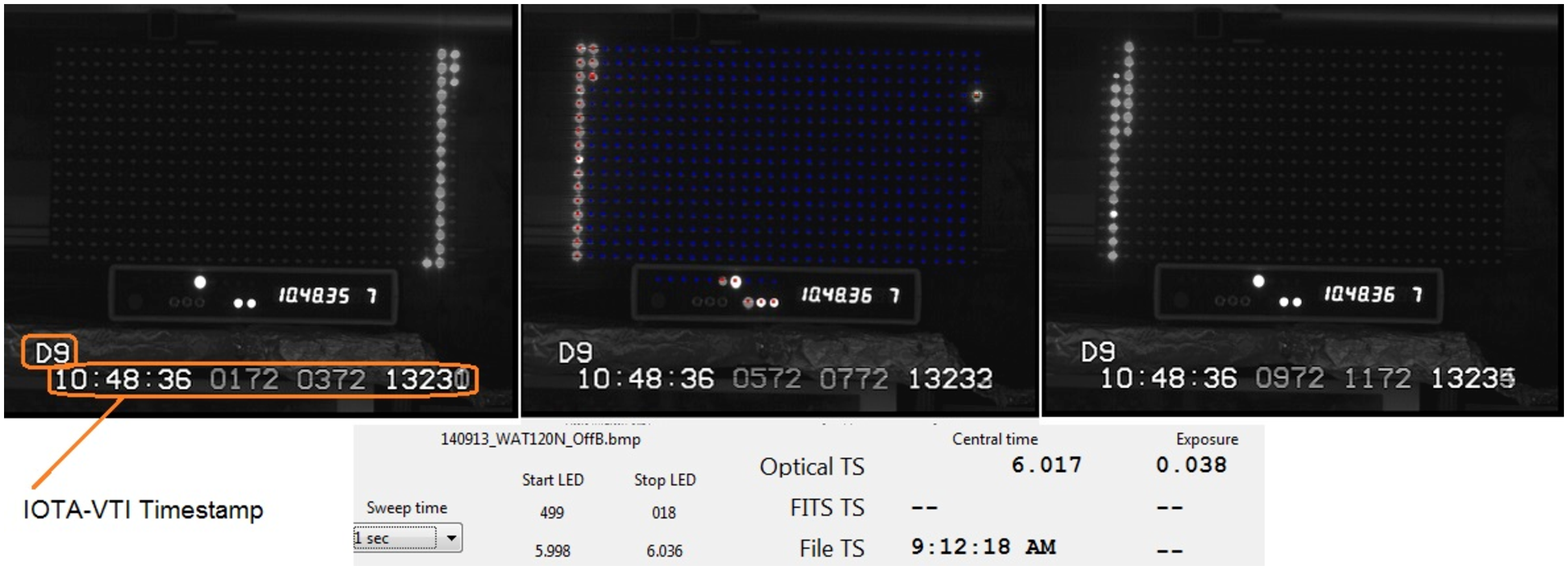}
\caption{Three frames from a GPS timestamped video occultation camera. For details see section~\ref{sec:test1}.}
\label{Fig3}
\end{center}
\end{figure*}

\section{TESTING OF A GPS BASED VIDEO CAMERA USED FOR OCCULTATIONS}
 \label{sec:test1}

The Watec 120N analog video camera has been used for occultation recordings of Pluto and main belt asteroids \cite{CharonOcc2008Paper}. The camera has the ability to synthesise long duration exposures by accumulating (stacking) short duration images in an internal buffer, then output the stacked images in accordance with the NTSC/PAL standard.

Figure~\ref{Fig3} shows three consecutive video frames from a PAL Watec 120N camera with no accumulation (i.e. 40~msec of imaging time per picture) while recording a view of SEXTA. The images were time-stamped with a commercially available video time inserter (IOTA-VTI, VideoTimers Inc, CA, USA); the VTI time stamp being shown at the bottom of each image, and circled on the left image in orange. The middle image has the SEXTA derived optical time stamp provided in the section below the image, along with the file creation time stamp.

The first item of note is that there is very little dead time between frames. The left picture ends with the 499$^{th}$ LED on the SEXTA panel being illuminated at the end of the fifth UT integer second. The middle picture shows the 500$^{th}$ LED of the fifth second illuminated and then 19 LEDs (38~msec) in the sixth second. The right picture shows the LED for the 38$^{th}$~msec partially illuminated, indicating that the dead time is much less than 2~msec.

The second item of note is that the middle of the illuminated LED band is twice as bright as the other illuminated LEDs. This is because the Watec camera records in an interlaced manner, with the even raster lines being exposed first, and the odd raster lines being exposed second (thus seeing different times, even though they are adjacent to each other on the image). The bright LED occurs where the second field begins exposure while the first field is still being exposed. The amount of field overlap or separation varies with exposure settings and must be determined for a given camera at a given setting.

The third item of note is that the SEXTA central image timestamp reads 40~msec before the IOTA-VTI timestamp. This is due to the delay induced by the buffer system of the camera and is the instrument delay (ID) time \cite{ProAmPaper} \cite{HarringtonPaper}. It is common with analog video integrating cameras, with the amount of ID varying between different devices and settings, but constant for a given device and setting. The ID must be subtracted from the IOTA-VTI timestamp to obtain the correct timestamp.

The issue of ID varying with analog camera settings is a long standing problem. No automated means of addressing it presently exists.  It remains a task for the operator to compile a table of ID for each camera setting, and then manually apply it where time values are needed.  

\section{TESTING OF NTP BASED CCD CAMERAS USED FOR OCCULTATIONS}
 \label{sec:test2}
\subsection{Preamble:-}
The SBIG (Santa Barbara Instruments Group) family of cameras have been used for TNO occultations on several occasions \cite{QuaoarOccPaper} \cite {CharonOcc2008Paper}. Unlike the video camera described above, theses cameras are driven by software running on a tethered PC; this controls the camera gain, initiates image acquisition, and defines the region of interest (ROI) on the CCD chip; and the camera downloads the ROI to storage on the computer.  The camera control software and host PC are therefore critical parts of the imaging system.

Image timestamps are derived from the PC system clock, which is synchronised to UT by means of the Network Time Protocol (NTP). NTP requires an active internet connection to operate, and makes use of NTP servers available on the internet, to determine  UT to a variable degree of error. The NTP system can, when connected to low stratum number NTP timeservers over a low latency network connection, offer PC system times which are within tens of milliseconds of true UT \cite{NTPseismoPaper}. 

\subsection{Methods:-}
We tested two SBIG CCD cameras; the ST10XE using CCDops v5.6 (Santa Barbara Instruments Group, CA, USA) as the control program, and the ST8 using both MaximDL v5.03 (Diffraction Limited, Ottawa, Canada) and CCDSoft v5.00.210 (Software Bisque, CO, USA) as the control programs. All cameras had mechanical shutters and a USB connection to the PC. 

\subsubsection{ST10XE + CCDops}
We installed the SBIG ST10XE camera + CCDops on a Windows~7 32-bit computer with a Core~I7 processor, 4~GB RAM, a 1~TB 5400~rpm hard drive, and provided with an ADSL2+ network connection of $\sim$1~Mbps.  NTP was synchronised using a human machine interface called Dimension4 (D4), freely available as a download, which allows the user to run NTP as a service on Windows 7 machines, synchronise to designated NTP servers, and maintain a log of the offsets from UT over time.  D4 was peered with a server from the Australian NTP Pool, and a log of the offsets was collected during the camera testing run. The offset time during the run was +115~msec, i.e the PC was ahead of UT by this amount.

The camera was set to take images with an exposure duration of 250~msec, and as frequently as the camera and PC software could work, which was an image about every 4~seconds. The imaging run was over 17~minutes (266 frames), which would be a reasonable period for a TNO occultation recording. 

The SEXTA panel was configured to have a sweep time of 5~seconds, giving a temporal resolution of 10~msec per LED illumination time.

\subsubsection{ST8 + MaximDL and CCDSoft}
We installed the SBIG ST8 camera, MaximDL and CCDSoft on a Windows~7 64~bit computer with a Core i7 processor, 8GB RAM, and provided NTP services via a LAN Stratum 3 NTP timeserver synced to two regional Stratum 1 timeservers (time.uwa.edu.au and dns.iinet.net.au).  NTP on the PC was synchronised using Tardis, a freely available interface with updates running every 60 seconds. 

The camera was set to take images of 1.9~second duration, and the SEXTA panel was configured for a sweep time of 2 seconds, giving a temporal resolution of 4~msec.  We took 100 images with each control program, then rotated the camera housing 180 degrees so that the CCD saw the SEXTA panel sweeping from right to left instead of the normal left-to-right, and repeated the 100 exposures.  This was to elucidate any effect that the mechanical shutter might have on the imaging exposure, as the shutter is not instantaneous in its operation but sweeps over the field always in one direction with respect to the CCD.

 \subsection{Results:-}
 \subsubsection{CCDops}
 \label{sec:test25}
The major finding was that the CCDops program wrote a timestamp to the header which resolved to the second, and no further. Thus, if an image was begun at 01h~23m~45.678s, the FITS header would be written as 01:23:45.000. This produced most of the error between optical timestamps and FITS. The FITS time was the image start time (which is what the FITS standard requires for the DATE-OBS field), rather than the image central time which would be what an astronomer would use in calculations.

\begin{figure*}[h!]
\begin{center}
\includegraphics[width=1.9\columnwidth]{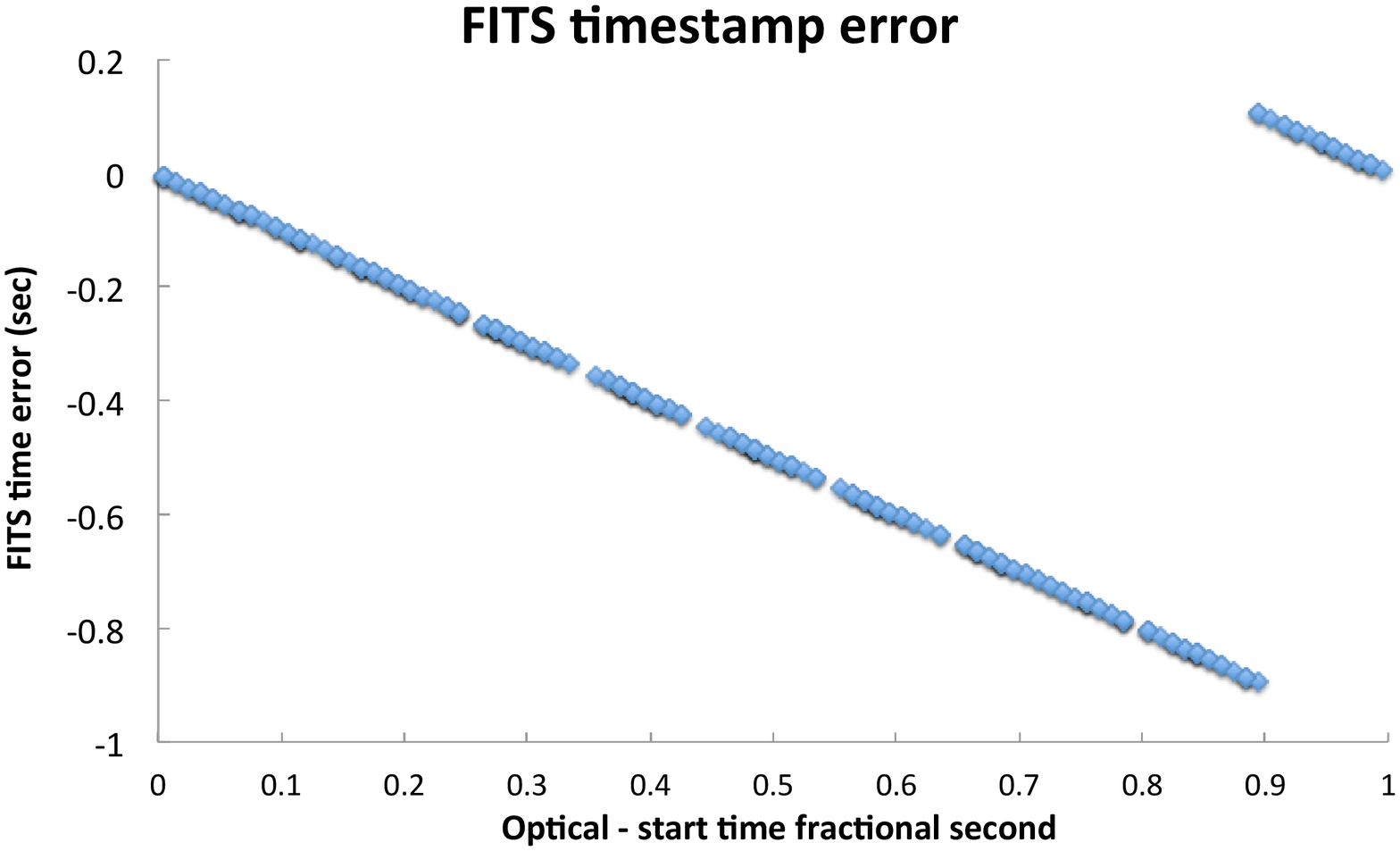}
\caption{Optical-to-FITS fractional second error showing NTP offset for an NTP based CCD camera + CCDops. }
 \label{Fig4}
\end{center}
\end{figure*}

Secondly, the PC clock was ahead of UT by 115~msec at the time of the imaging run, as indicated by D4. This is a high offset for an NTP synced computer, and a more reasonable result would be around 20~msec. Possible reasons were the short time that D4/NTP was running on the computer (about 4~hours before the imaging run) which is known to cause larger offsets \cite{NTPseismoPaper}. We graphed the time (see Figure~\ref{Fig4}) within any UT second when an image was started (as measured by optical time stamp). We measured the error between FITS and optical start time, which should have been between zero and one second (due to the integer second resolution of the FITS timestamp) if the PC was synchronised perfectly with UT. We found that the knee in the graph occurred at the optical fraction second time around 875 to 895~msec, rather than 1000~msec. The disparity of 105 to 125~msec is in good agreement with the D4-reported offset of +115~msec.

\begin{figure*}[h!]
\begin{center}
\includegraphics[width=1.9\columnwidth]{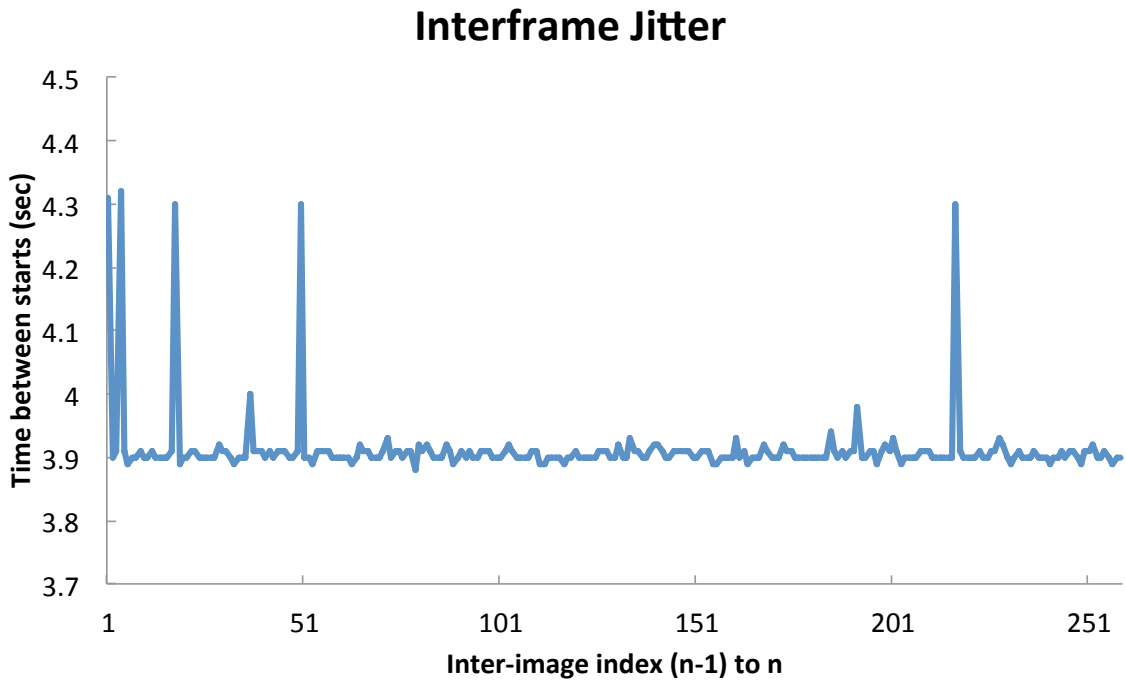}
\caption{Imaging cadence of an NTP based CCD camera + CCDops. }
 \label{Fig5}
\end{center}
\end{figure*}
\begin{figure*}[h!]
\begin{center}
\includegraphics[width=17.5cm]{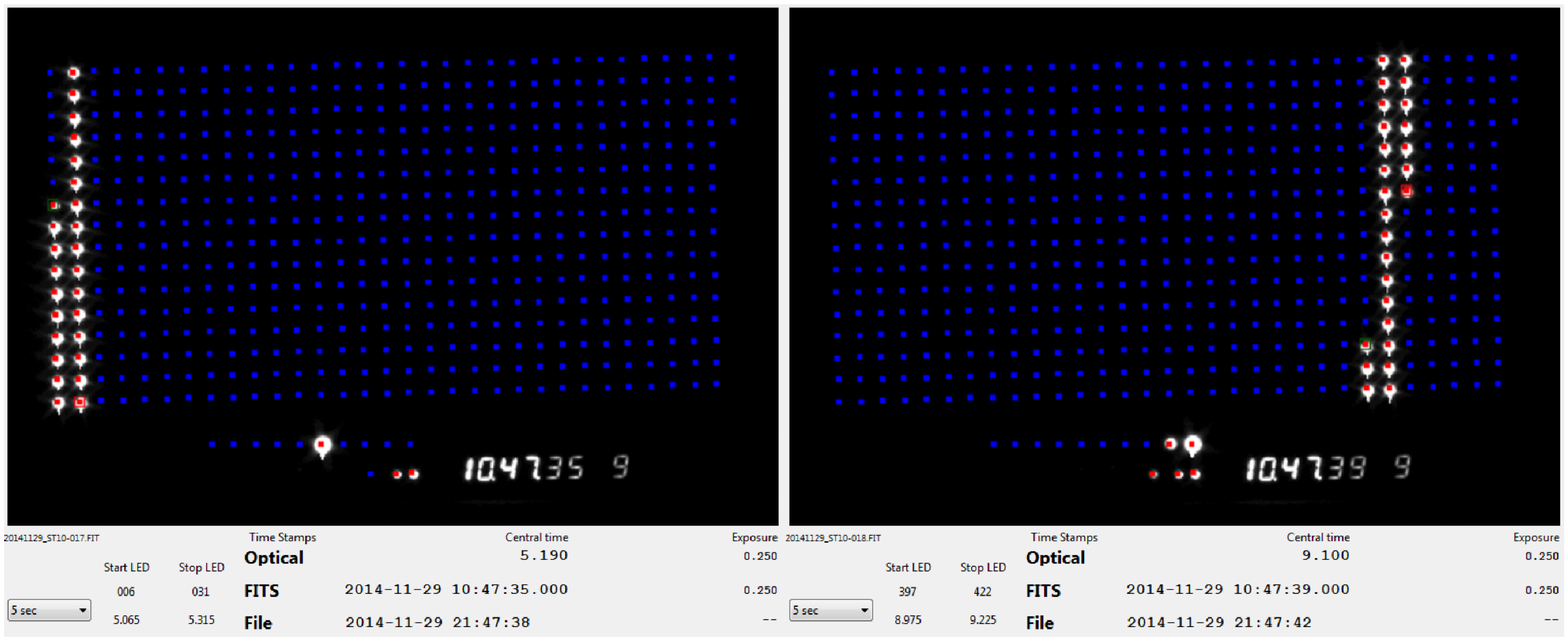}
\caption{Two consecutive frames from an NTP timestamped CCD occultation camera with CCDops. For details see section~\ref{sec:test25}.}
\label{Fig6}
\end{center}
\end{figure*}

Thirdly, imaging cadence (see Figure~\ref{Fig5}) was not particularly steady. The average image cadence was 3.91~sec, with a jitter of -32 to +408~msec. This jitter would be difficult or impossible to detect using FITS information as it presently stands. 

The FITS header exposure times (image duration) were very consistent and agreed well with the optical information. The maximum and minimum exposure was 240~msec and 260~msec, with the mean and standard deviation being 250.9~msec$\pm$0.0034~msec.

Imaging time to download time was 0.25 vs 3.91~seconds, which is acceptable for a testing regime. See Figure~\ref{Fig6}.

\subsubsection{MaximDL and CCDSoft}
\label{sec:test3}
The MaximDL program wrote timestamps resolved to the centisecond, while CCDSoft wrote millisecond resolution timestamps to the FITS header, but the delay between header start time and optical start time for MaximDL was 802.2~msec average, with excursions of $\pm$20~msec; while with CCDsoft, the delay was much less, being 79.1~msec average, with excursions of $\pm$17~msec. This represents an order of magnitude improvement in timestamp accuracy with CCDSoft. 

Image cadence for MaximDL and the ST8 with a 1.9 sec exposure was 9.8535~sec average, with $\pm$54~msec jitter.  CCDSoft and the ST8 had a very similar cadence of 9.9893~sec average, and an identical jitter of $\pm$54~msec.

Image duration was identical for both MaximDL and CCDSoft, with an average of 1.92~sec $\pm$18~msec. 

\begin{figure*}[h!]
\begin{center}
\includegraphics[width=17.5cm]{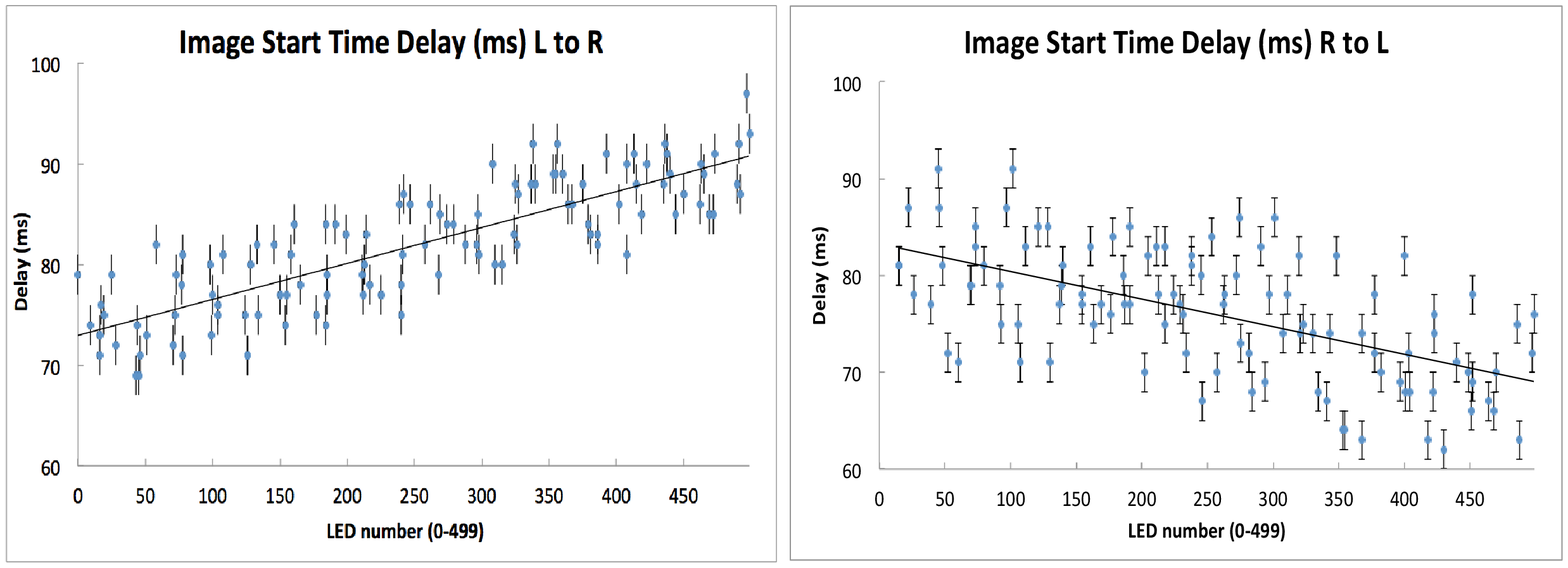}
\caption{CCDSoft timestamp delays shown with respect to SEXTA panel LED number. For details see section~\ref{sec:test3}.}
\label{Fig7}
\end{center}
\end{figure*}

The mechanical shutter introduced a small (but measurable) left-to-right time bias across the CCD of around 14~msec. That is, an event recorded near the leading edge of the CCD (which opens to light first, and which we consider here as the left edge) is delayed less than an event recorded near the trailing edge of the CCD (which opens to light only when the shutter has traversed the CCD).  This was confirmed when the camera was rotated 180 degrees with respect to the SEXTA panel, so that the SEXTA sweep was from right-to-left.  See Figure~\ref{Fig7}.

 \subsection{Discussion:-}
The three CCD camera control software programs examined produced timestamps with widely varying fidelities to UT.  The worst case was CCDops, with a delay of 1~second from true, due to integer second time recordings, and a cadence jitter of 400~msec.  Add to this an unknown NTP offset and it is easy to appreciate the difficulties experienced by the Chariklo researchers mentioned in Section 2.

The best case was CCDsoft, with a delay of less than 100~msec from true, and a cadence jitter of 40~msec - an order of magnitude improvement over the worst case.  The NTP offset  remains an undocumented quantity. 

The image duration variability for all programs was $\pm$18~msec, and we speculate that this may be due to the exposure being timed by the hardware in the camera rather than the host PC.

The most severe timing issue was the fact that the NTP offset was not recorded in the FITS header by any of the programs tested. Because of this, some other means must be employed to verify that NTP is operational and has reasonably low offsets. If this is not done, the FITS timestamp would be in error from true by an unknown number of seconds.

The imaging cadence variation is less amenable to simple fixes, and may depend on what the host computer is doing (i.e. other housekeeping tasks). This topic is beyond the scope of this discussion. 

\subsection{Limitations:-}
This examination of two NTP based cameras with three commercial programs is a good beginning, but cannot be considered exhaustive testing of any system. It is entirely possible that further testing may uncover outlier events which compound any errors detected here by orders of magnitude.

\section{INFERENCES FROM TESTING}

The SEXTA result for a given camera and recorder system does not necessarily provide assurance that the camera system will continue to perform in the same way in the future. Such assurance would come from repeated testing over some reasonable period of time.

The Watec analog video camera and GPS-based video time inserter examined here have been found to be stable and consistent in behaviour. This offers confidence that results obtained in the future can be relied upon.

The SBIG cameras and NTP-based time references examined here have more variable results which could compromise occultation recording timings to the extent seen in the Chariklo occultation mentioned in Section~\ref{sec:motivation}. Some avenues of exploration remain to improve the method.

\section{CONCLUSION}

A system for verifying time-stamped image time and duration, to 2~msec precision and within 1~msec of GPS fiducial time, is described. The system is very low cost and requires minimal assembly. Parts are readily obtainable. Source code and wiring diagrams and a built app with source code for analysing the image time stamps are provided and available for download. 

\section*{SUPPLEMENTARY DATA}
The supplementary data with additional documentation, web links for more references, build notes, source code, and applications for Windows~7+, MacOS~10.7+, and Linux Ubuntu~12.04 can be found at: 
\newline
\newline
http://www.kuriwaobservatory.com/SEXTA/
\newline
http://www.tonybarry.net/

\section*{ACKNOWLEDGEMENTS}
The original anecdotal work that inspired the development of SEXTA can be found at:-
\newline
\indent http://www.dangl.at/exta/exta\_e.htm
\newline
The author, Gerhard Dangl, has stated :-
\begin{quotation}

``...Because of the relative complex design required for this functions and the big display with a large number of LEDs the goal was not to develop this device for reproduction. So at the moment it will stay as a very useful prototype device.'' (Page dated October 10, 2012, retrieved on 2014-07-24 08:41:00 UT) 
\end{quotation}
 
As such, this work cannot be tested or verified, or used by others for further investigation; and this absence led to the development of SEXTA. 
\newline
\newline
The authors would also like to thank Mr. Edward Dobosz of the Western Sydney Amateur Astronomy Group for assistance with the testing of an SBIG camera. 

\section*{FINANCIAL SUPPORT}
T.B. was supported by a Joint Research Engagement (JRE) grant from the University of Sydney (2012 -- 2014).

\section*{CONFLICTS OF INTEREST}
None.


\end{document}